\documentclass[twocolumn,twoside,slac_two]{revtex4}
\usepackage{graphicx}
\usepackage{fancyhdr}

\begin{document}

\title{Probing the Heavy Flavor Content in $\boldmath t \bar{t}$ Events and Using $t \bar{t}$ Events as a Calibration Tool at CMS}

%

\author{Roberta Volpe on the behalf of CMS collaboration}
\affiliation{Universita' degi Studi and INFN Perugia, via A. Pascoli 1, 06123 Perugia, Italy }
\begin{abstract}
We present two analyses dedicated to measure the ratio of branching ratios of the top quark, $R=B(t \rightarrow W b )/(t \rightarrow W q )$ ( where $q=d,s,b$), using ttbar events with either one or two prompt isolated leptons (e or mu) in the final state. 
Furthermore the framework of the dileptonic analysis was used also for a feasibility study of the measurement of b-tagging efficiency, by assuming the $R$ value to be the Standard Model one.
Data-driven techniques to control the background in the selected events are discussed and the expected simulation results are presented.
\end{abstract}
\maketitle
\thispagestyle{fancy}
\section{Introduction}
\label{sec:intro}
Top quarks decay mostly to $Wb$, while the final states $Wd$ and $Ws$ are suppressed by the square of the CKM matrix elements $|V_{td}|$ and $|V_{ts}|$.
Besides single top studies, $|V_{tb}|$ can be obtained also through top pairs production, by measuring $R =B(t \rightarrow W b )/(t \rightarrow W q) $, with $q=d,s,b$, and assuming that exactly 3 generations of quarks exist, as the Standard Model (SM) predicts;
indeed,
by imposing the unitarity of the $3 \times 3$ CKM matrix, such ratio is $R = |V_{tb}|^2/(|V_{td}|^2+|V_{ts}|^2+|V_{tb}|^2) =|V_{tb}|^2 $.
Without any assumption on the number of generations of quarks, an $R$ measurement is still useful to put constraints on $V_{tb}$ and, more importantly, it can give a clue on the existence of a fourth generation; indeed in such scenario, $R$ is appreciably less than the SM value \cite{Alwall:2006bx}. 
The most recent $R$ measurement obtained by CDF with $\sim$ 162 pb$^{-1}$ is $R>0.61$ at 95 $\%$ C.L.~\cite{Acosta:2005hr};
D\O~ measured $R$ simultaneously with the $t \bar{t}$ cross section and obtained the value $R=0.97_{-0.-08}^{+0.09}$
and a limit $R>0.79$ at 95 $\%$ C.L. with $\sim$ 900 pb$^{-1}$~\cite{Abazov:2008yn}.
The direct measurement of the CKM element $|V_{tb}|$ (predicted by the SM as $|V_{tb}|=0.999133_{-0.000043}^{+0.000044}$)~\cite{Amsler:2008zzb})
is possible only by means of the study of single top production
and currently the only available
measurements are from D\O~\cite{Abazov:2009ii} and CDF \cite{Aaltonen:2009jj} experiments.
In the CMS experiment \cite{cms}, two feasibility studies of the $R$ measurement have been carried on, one using selected semileptonic $t \bar{t}$ events \cite{PASsemil} and described in Sec.~\ref{sec:semilep}, the other using selected dileptonic $t \bar{t}$ events \cite{PASdilep} as described in Sec.~\ref{sec:dilep}.
Both the analysis use data-driven methods in order to estimate the irreducible background contribution and consider the number of b-tagged jets as the physical observable, therefore the b-tagging efficiency must be fixed to a value obtained from an independent measurement.
Furthermore, 
the framework of the analysis can be used also to the aim to perform a measurement of b-tagging efficiency by assuming the $R$ value to be the SM one; such study was performed for the dilepton channel and is described in Sec.
\ref{subsub:effbmeasurement}.\\  
\section{General Method}
\label{sec:method}
The parameter $R = B(t\rightarrow Wb)/B(t \rightarrow Wq)$ is measured by 
counting the number of jets originating from $b$-quark ($b$-jets) in $t \bar t$ events. 
The number of $b$-tagged jets 
depends, beyond the $R$ value itself, on the $b$-tagging efficiency ($\epsilon_{b}$) and the
mis-tagging probability ($\epsilon_{q}$).
Therefore, the probability to have a given number $i$ of 
$b$-tagged jets 
is a function of $R$, $\epsilon_{b}$ and $\epsilon_{q}$. It is called $\varepsilon_i(R;\epsilon_{b},\epsilon_{q})$ and can be expressed 
as follows: 
\begin{eqnarray}
\varepsilon_i(R;\epsilon_{b},\epsilon_{q}) =  R^2 P_i(t \bar{t}\rightarrow bWbW) \nonumber \\
 + 2R(1-R) P_i(t \bar{t}\rightarrow bWqW)\nonumber \\
 + (1-R)^2 P_i(t \bar{t}\rightarrow qWqW)   
\label{eq:epsilon}
\end{eqnarray}
where $q$ can represent an $s$ or $d$ quark and each $P_i$ (probability for
a definite $t\bar{t}$ decay of having $i$ $b$-tagged jets in the final
state) depends on $\epsilon_{b}$ and $\epsilon_{q}$.
This function is used to fit the distribution of the number of $b$-tagged jets ($n_{btag}$) to measure the value of the $R$ parameter.
In order to identify the flavor of the jets, specific algorithms are used. 
For this study, the {\it Track Counting} (TC) and {\it Jet Probability} (JP)~\cite{PTDRVOL1}
algorithms are used to tag the b-jets.
The efficiency of the TC and JP algorithms can be measured
in QCD events with reconstructed jets containing muons.
The $p_{Trel}$ method~\cite{BTV_07_001} exploits the distribution of the relative transverse momentum of the muon with respect to the 
jet to estimate the number of $b$~jets present in data.

\section{Semi-leptonic $\boldmath t \bar{t}$ analysis}
\label{sec:semilep}
The final state of the semi-leptonic $t \bar{t}$ decay channel
(one $W \rightarrow q \bar{q}^{\prime} $ and the other $W \rightarrow l \nu_l $ )
is characterized by two quarks coming from the direct decay of top quarks, two quarks coming from 
the decay of one $W$ and a lepton and a neutrino from the other $W$ decay.
Therefore the final experimental signature is four or more jets, 
a single lepton (electron or muon) and missing transverse energy.
The generation of Monte Carlo signal and background samples is described in \cite{PASsemil}.
The following results refer to an integrated luminosity of 1 fb$^{-1}$.
\subsection{Selection and Event Reconstruction}
\label{subsec:selesemil}
The selection starts with the High Level Trigger (HLT) requests: a lepton with enough large $p_{\rm{T}}$ ($p_{\rm{T}}>15$ GeV for muons or $p_{\rm{T}}>18$ GeV for electrons).
The details of the physics objects reconstruction are in \cite{PASsemil} and references therein.
Offline electron reconstruction and identification 
is performed by using tracker and electromagnetic calorimeter information and the muon reconstruction uses both tracker and muon chambers sub-detectors information.
An isolation variable for the leptons is defined
as 
the ratio between the sum of $p_{\rm{T}}$ of the tracks and energies of calorimetric 
deposits around the candidate and the $p_{\rm{T}}$ of the lepton candidate itself.
The lepton candidate must have such isolation variable less than 0.1 and $p_{\rm{T}}>30$ GeV/c.
If more than one lepton is selected, the event is rejected.
The jet reconstruction algorithm  
uses the calorimetric energy deposits with a 
seed threshold of $E=1$ GeV and performs an 
iterative cone procedure with radius $\Delta R = \sqrt{\Delta \phi^2 + \Delta \eta^2}=0.5$.
The jet candidates are selected by requiring $E_{\rm{T}}>40$ GeV and $|\eta|<2.4$; 
in order to reject 
fake jets, they are required to have
the fraction of electromagnetic energy to the total energy
less than one
and to be far enough from the lepton candidate ($lep$) 
by imposing 
$\Delta R(jet,lep) > 0.5$. 
The missing transverse energy ($\displaystyle{\not}{E}_{T}$) used in this analysis is 
computed by performing the vectorial sum of the energy deposits in the calorimeters. 
The reconstruction of neutrino momentum is needed to compute the leptonic 
top quark mass; 
the transverse component comes from the $\displaystyle{\not}{E}_{T}$ value 
while the longitudinal component is determined from the four-momentum
conservation of the $W$ boson decay.
A useful kinematic variable to reduce the background contamination is Centrality.
It represents the fraction of the hard scattering going in the transverse plane and it is defined as:
\begin{equation}
  \label{centr}
  \rm{Centrality} = \frac{\sum E_{\rm{T}}} {\sqrt{\Big(\sum E\Big)^2-\Big(\sum p_z\Big)^2}}
\end{equation}
It is required to be larger than 0.35.
The final step of the event reconstruction is the computation of the invariant 
masses using the selected reconstructed objects.
Among the selected jets, the four with largest $E_{\rm{T}}$ are considered as 
coming from the decays of the two top quarks and of the hadronic $W$. 
While the selected lepton and the missing energy are known to come from a $W$ decay, 
the assignment of the four chosen jets to the partons       
has to be determined. 
In order to choose the right combination, a two step association is used.
Beforehand the masses and the widths of the hadronic $W$ boson and the tops are 
obtained from simulation.
The distributions of the three invariant masses of the reconstructed objects
 well matched to the generated particles are used to obtain the parameters 
$m_{Whad}$, $m_{tHad}$, $m_{tLep}$, $\sigma(m_{Whad})$, $\sigma(m_{tHad})$ and $\sigma(m_{tLep})$.
First the hadronic $W$ boson is reconstructed by computing the invariant mass of every pairs of jets among the four. 
The pair with the nearest invariant mass to the $W$ one, namely $ij$,
 is chosen. 
The following cut, is applied:
\begin{equation}
  |m_{ij}-m_{Whad}|< \sigma(m_{Whad}) 
\end{equation}
The second step is the association of the two remaining jets ($k$ and $p$) to the partons coming from the direct 
decay of top quarks.  To this end a $\chi^2$ based on the two top quarks masses is defined:
\begin{equation}
 \label{eq:chi2}
  \chi^2 = \Bigg( \frac{m_{ijk}-m_{tHad}}{\sigma(m_{tHad})} \Bigg)^2 
+ \Bigg( \frac{m_{l \nu p}-m_{tLep}}{\sigma(m_{tLep})} \Bigg)^2 
\end{equation}
where $i$ and $j$ are the 2 jets chosen as coming from the $W$ boson decay. 
Now the only combinatorial ambiguity lies in the choice of which one 
of the two remaining jets is associated to which of the two top quark.
The association that minimizes the $\chi^2$ is assumed to be the correct one. 
We consider the events with a large $\chi^2_{min}$ as events which are wrongly 
reconstructed, so the cut $\chi^2_{min}< 4$ is applied.\\
After the whole selection, the expected event number with an integrated luminosity 
of 1 fb$^{-1}$ is 2650 for semileptonic 
$t \bar{t}$, while the main background processes are: 109 for other $t \bar{t}$, 
260 for $W+jets$, 52 for $Z+jets$, 52 for $tW$ and 56 for QCD di-jet. 
The $b$-tagging algorithm adopted in this analysis is the $Jet Probability$  
\cite{PTDRVOL1}
and the chosen working point is such that
$\epsilon_{b}=(79 \pm 1)$\% and $\epsilon_{q}=(13 \pm 1)$\%.
\subsection{Background Subtraction}
\label{subsec:bkgSubSemil}
\begin{figure}[h]
\centering
\includegraphics[angle=90,width=90mm]{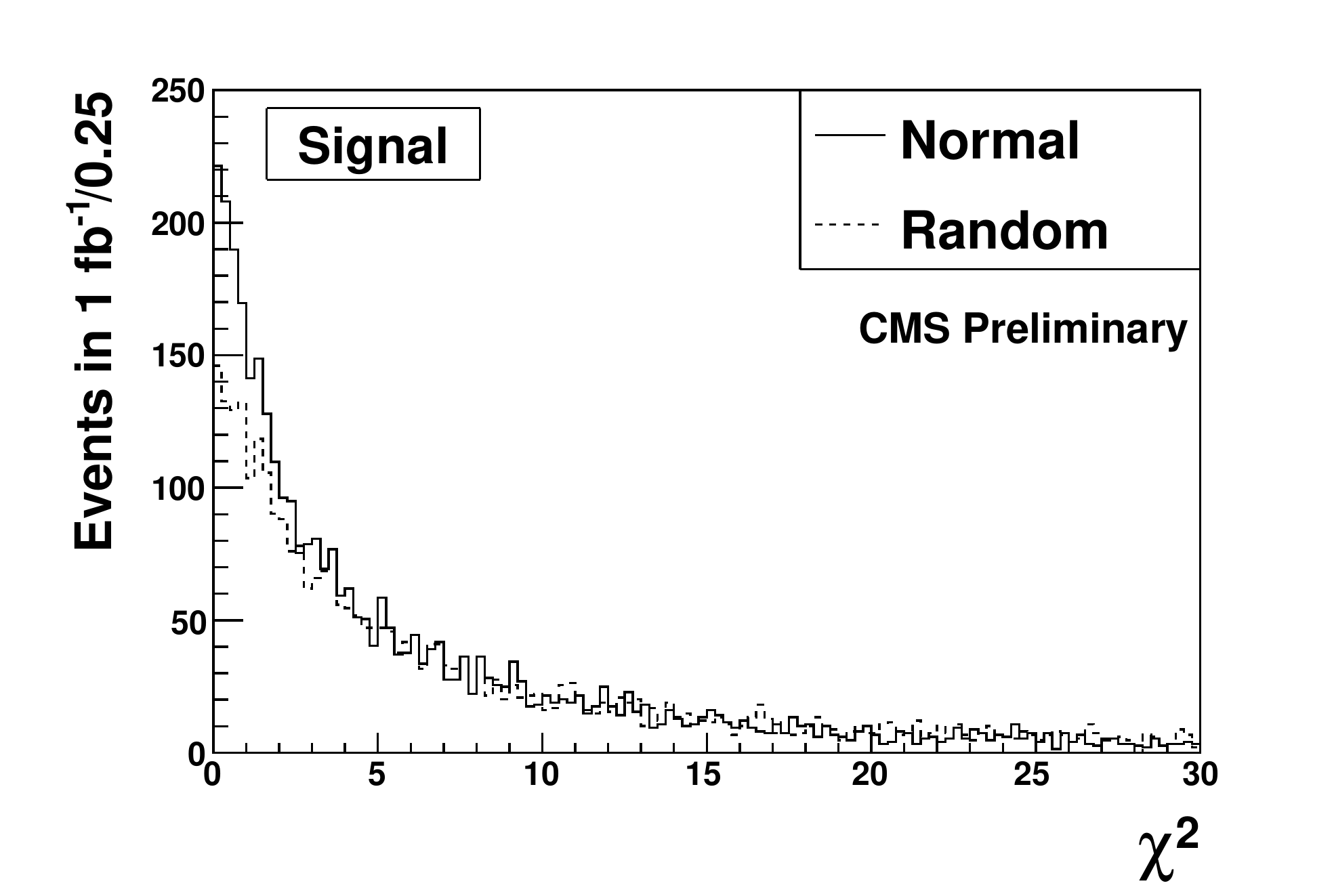} \\
\includegraphics[angle=90,width=90mm]{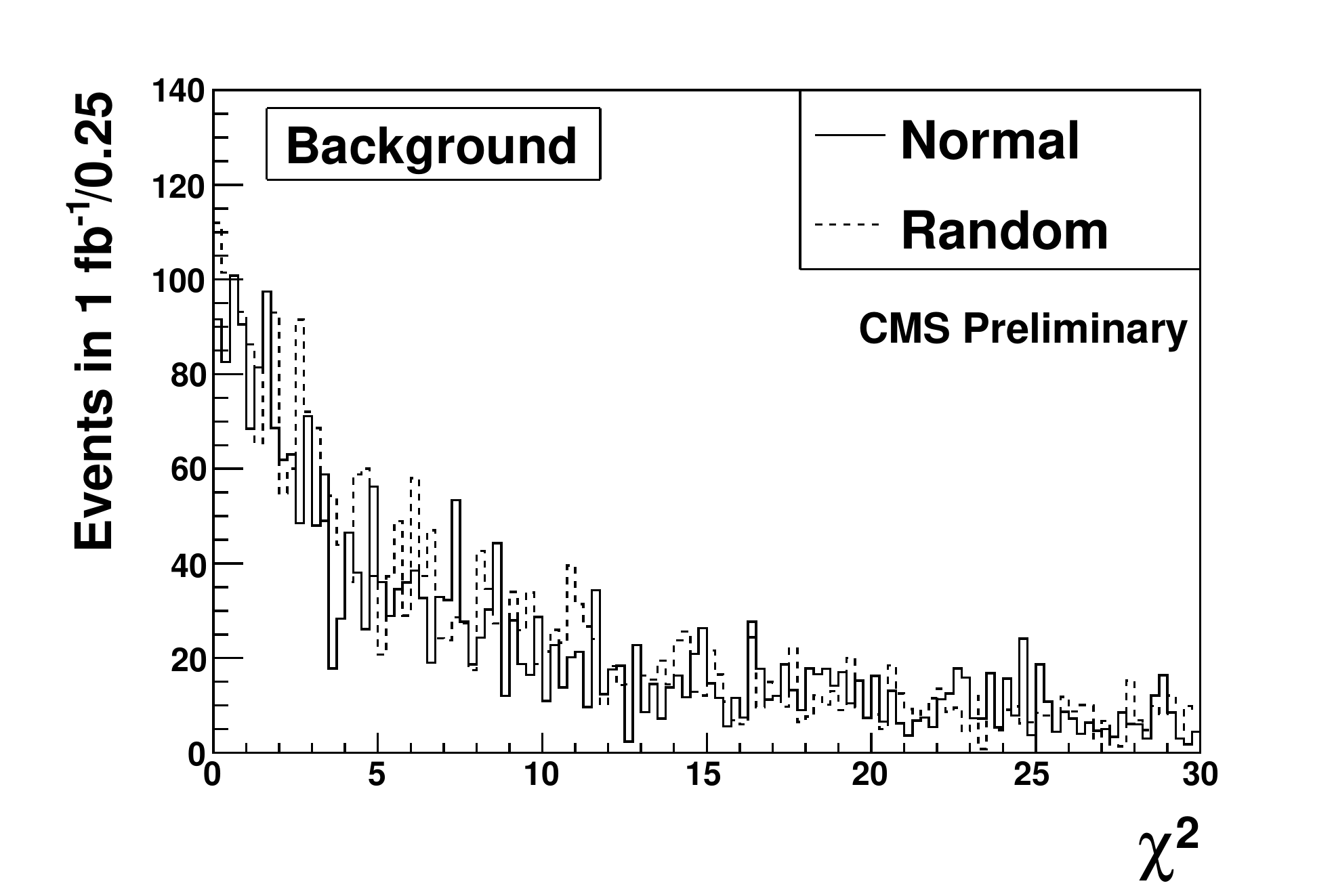} 
\caption{Up: $\chi^2_{min}$ distribution of the $signal$ sample (as defined in the text).
 Down: $\chi^2_{min}$ distribution of the complete background sample. 
Both the figures show the $\chi^2_{normal}$ (solid) and the $\chi^2_{random}$ (dashed) distributions.}\label{fig:chi2_sigback}
\end{figure}

\begin{figure}[h]
\centering
\includegraphics[angle=90,width=90mm]{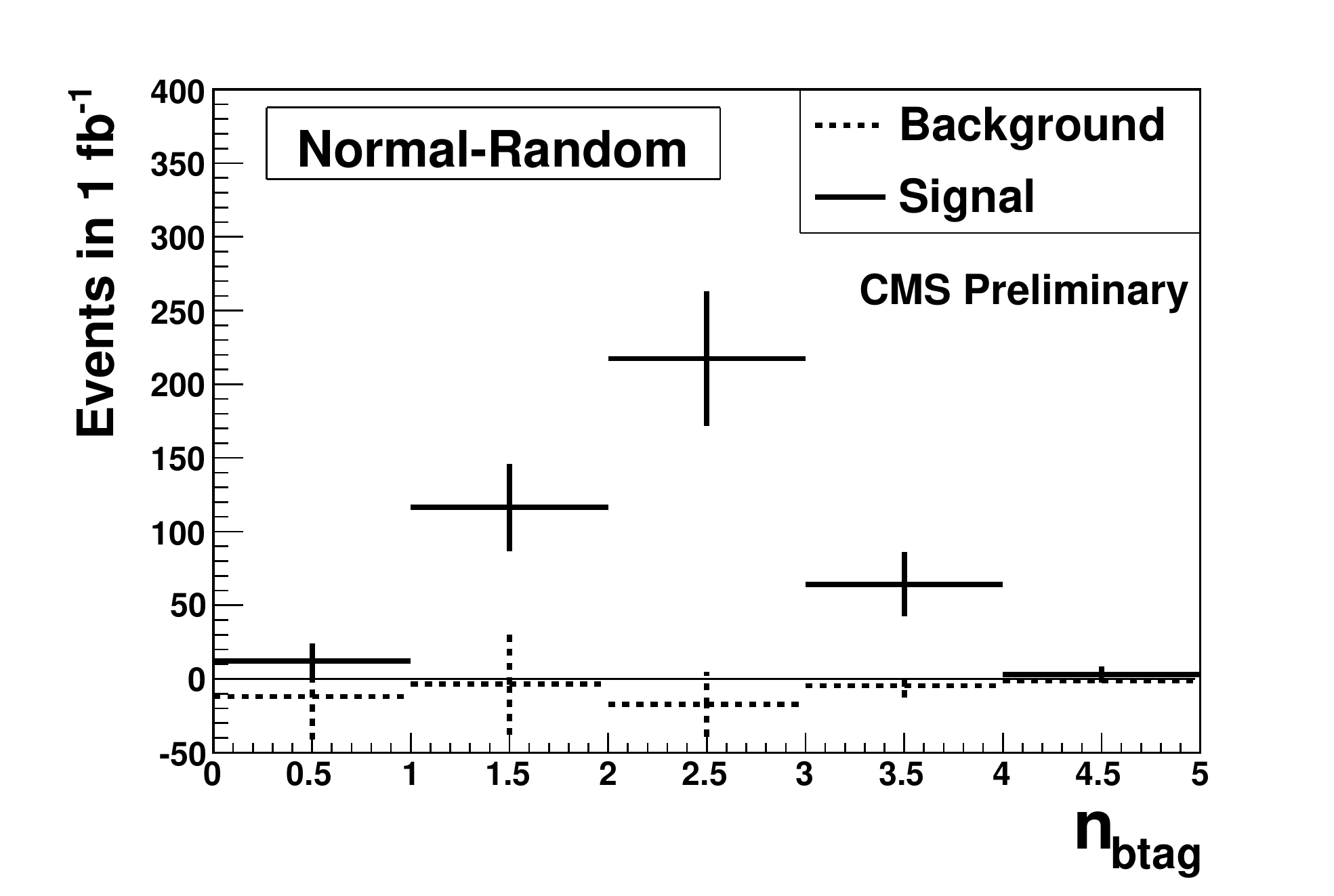} \\
\includegraphics[angle=90,width=90mm]{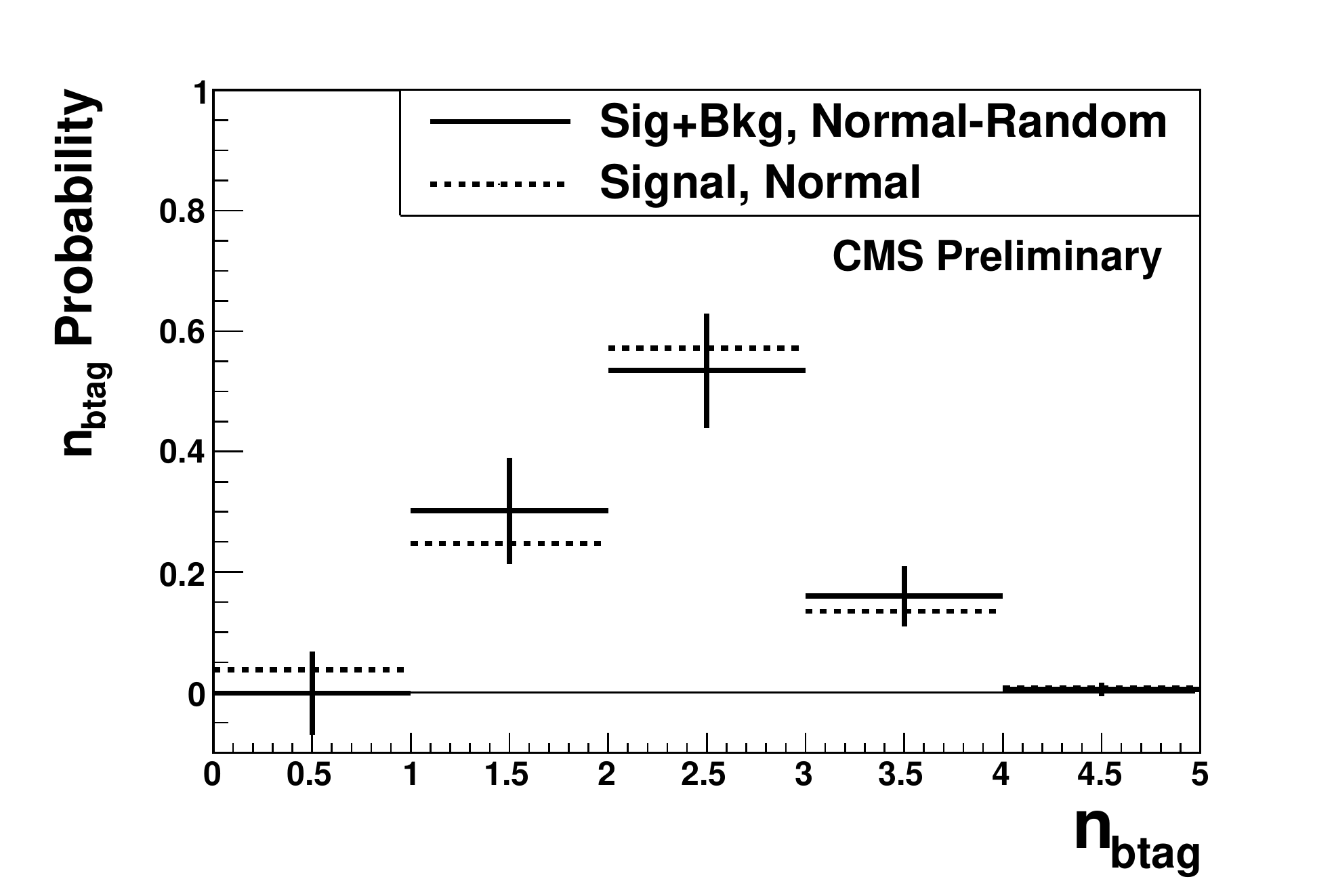} 
\caption{
Up: $n_{btag}^{normal}$-$n_{btag}^{random}$ distribution for $signal$ (solid) and $background$ (dashed) events normalized to $L=1$ fb$^{-1}$ . 
Down: $n_{btag}^{normal}$-$n_{btag}^{random}$ distribution for the whole data sample (solid) and
 $n_{btag}^{normal}$ distribution of the only $signal$ (dashed) normalized to unity. 
}\label{fig:btagsub}
\end{figure}

The $\chi_{min}^2$ defined in Eq.~\ref{eq:chi2}, and referred to as $\chi^2_{normal}$ 
in the following, has a peak at low values of $\chi^2$ for correctly reconstructed semileptonic $t \bar{t}$ events, called
$signal$ in the following. 
Background and incorrectly reconstructed $t \bar{t}$ events 
($Background$ in the following) lead
to low values of $\chi^2_{normal}$ only due to random combinatorics.
Therefore if the direction of one of the selected jets is artificially changed, 
the mass $\chi^2$ distribution should remain the same for background events, 
while we expect the distribution for $signal$ events will appreciably change.
We can define a $\chi^2_{random}$ just like the $\chi^2_{normal}$, 
but computed by assigning a 
random direction to one of the two jets considered as coming directly from the tops.
We decided to change the direction of the one with highest transverse energy. 
Uniform distributions for $\phi$ and $\eta$ have been generated, allowing for $\phi$ in 
the range $(-\pi, \pi)$ and $\eta$ in the range $(-2.4, 2.4)$, 
as for the selected true jets. 
Then the $\chi^2$ procedure was repeated leading up to the new combination that gives 
the minimum $\chi^2$, called $\chi^2_{random}$.
Fig.~\ref{fig:chi2_sigback} shows the distribution of the $\chi^2_{min}$ variable 
separately for $signal$ and background events.
The $n_{btag}$ distribution of the events selected after the cut 
$\chi^2_{normal}<4$ 
will be referred as $n_{btag}^{normal}$ while the events selected after the cut 
$\chi^2_{random}<4$ will be referred as $n_{btag}^{random}$; 
Fig.\ref{fig:btagsub} (Upper panel) shows the result of the 
$n_{btag}^{normal}$-$n_{btag}^{random}$ 
subtraction for the $signal$ sample (solid) and for the background sample (dashed), the latter is 
compatible with a flat zero distribution.
Therefore, it is clear that if one considers the whole data sample, 
containing signal and background events,
and computes bin-by-bin the difference of the $Normal$ and $Random$ 
distributions, the resulting $n_{btag}$ distribution will be proportional to the 
distribution of the $signal$ only, as Fig.~\ref{fig:btagsub} (down) shows.
\subsection{Fit Results}
\label{subsec:fitSemil}
The distribution resulting from the bin-by-bin subtraction of the whole data sample, 
after normalization, is to be fitted with Eq.~\ref{eq:epsilon}.
In order to check the effectiveness of the method the fit was repeated assuming  
several $R$ values.  
Different values of $R$ ($R_{gen}$) were generated in the range [0.9, 1] by properly
weighting three samples where the decay of $t \bar t$ was forced: $t \bar t \rightarrow WbWb$, 
$t \bar t \rightarrow WbWq$, 
$t \bar t \rightarrow WqWq$.
The statistical uncertainty remains steady in all the range and it is 
$\sigma_{R}(stat)=0.12$. 
The measured values of $R$ agree within the statistical uncertainty with $R_{gen}$ in the
range $R_{gen}=[0.9,1]$.
\subsection{Systematic Uncertainties}
\label{subsec:errSemil}
The various uncertainties were estimated based on the anticipated knowledge of the 
CMS experiment after 1 fb$^{-1}$ 
of integrated luminosity \cite{PASsemil}. 
All systematic contributions were assumed to be uncorrelated, 
therefore the total systematic uncertainty has been computed by square summing. 
In order to check the impact of $\epsilon_b$ on $R$ measurement, its 
value was varied by 4\%.
Since the number of $b$-tagged jets from Eq.~\ref{eq:epsilon} does not take into account the
presence of $b$-jets from 
radiation, while the value of $\epsilon_b$ measured in real data~\cite{NotaCMS_B} does, a contribution due to such bias
is considered.
The systematic uncertainty associated to the jet energy scale is estimated by
shifting the calibrated transverse energy for each jet used in the analysis by a relative 5\%.
The effect of the difference in the selection efficiency between $t \bar t \rightarrow WbWb$ ($\varepsilon_{bb}$) and
$t \bar t \rightarrow WqWq$  ($\varepsilon_{qq}$) was estimated by 
varying $\epsilon_{bb}$ by $\varepsilon_{bb}-\varepsilon_{qq}$ (=0.04\%).
The $\chi^2$ cut was varied by $\pm 0.5$. 
The systematics study results for each source are summarized in 
Table~\ref{tab:systerr} together with the total value.
\begin{table}[h]
\begin{center}
\caption{Contributions to systematic uncertainty.}
\begin{tabular}{|l|c|}
\hline \textbf{systematics} & \textbf{$\sigma_{sys}$}
\\
\hline
\hline  b tagging efficiency & 0.04 \\
\hline b tagging efficiency  bias & 0.04 \\
\hline Jet Energy Scale & 0.09 \\
\hline  $\chi^2$ cut & 0.02 \\
\hline Selection efficiency   & 0.006 \\
\hline
\hline total  & 0.11 \\
\hline 
\end{tabular}
\label{tab:systerr}
\end{center}
\end{table}
\section{Dileptonic $ \boldmath t \bar{t}$ analysis}
\label{sec:dilep}
This study considers $t \bar{t}$ events were both the $W$ decay to leptons, the final state with an electron and a muon was chosen, as it is the channel with the largest cross section and smallest background.
The generation of Monte Carlo signal and background samples is described in \cite{PASdilep}.
All the results refer to an integrated luminosity of 250pb$^{-1}$. 
\subsection{Event Selection}
\label{subsec:seleDilep}
The event selection is tuned to identify leptonic final states with two prompt, 
isolated leptons with high transverse momenta
in the CMS detector.
The selection is detailed in \cite{PASdilep}.
 Data samples are triggered by requiring a non-isolated single muon ($p_T>9$~GeV/c) 
or a single electron ($E_{T}>15$~GeV).
Lepton candidates are reconstructed with $p_{T}\geq 20$~GeV/c 
in the fiducial region $\vert\eta\vert\leq2.4$ of the detector
and must satisfy identification and isolation requirements.
The leptons are required to be separated by $\Delta R>0.1$.
In the case of multiple selected leptons, the ambiguity is resolved by selecting $e\mu$ 
candidates with opposite electric charge and highest transverse momenta.
Jets are reconstructed using the seedless infrared-safe cone algorithm
and are required to have at least two calorimeter towers with a minimum $E_T$ 
sum of 2~GeV. Jets are required to have at least one assigned track so that the $b$-tagging algorithms 
can be applied. 
These cuts define the ``taggability'' requirements.
The energy of the jets is corrected for the $\eta$-dependence and absolute
$E_T$ using MC based corrections for generator level jets.
Taggable jets are selected with $E_T$(corrected)$\geq$~30~GeV/c and $\vert\eta\vert\leq2.4$.
Jet candidates are further required to be separated from the selected leptons by 
$\Delta R({\rm jet,lepton})\geq 0.3$ and to have an electromagnetic fraction EMF$<$0.98.
The total missing transverse energy, $\displaystyle{\not}{E}_{T}$, 
is corrected for the energy deposited by muons and it is required to 
be above $30$~GeV.
With 250~pb$^{-1}$ of integrated luminosity, after the described selection, 
the expected event number is 787 for dileptonic $t \bar{t}$, and the main background 
contributions are due to other $t \bar{t}$ (14 events),single top (29 events), 
Di-boson (10.5 events), $W/Z +jets$ (26 events). Therefore after the selection a signal to backround ratio of approximately 10 is expected. 
\subsection{The jet misassignment estimate}
\label{subsec:jetMisass}
\begin{figure}[h]
\centering
\includegraphics[height=80mm]{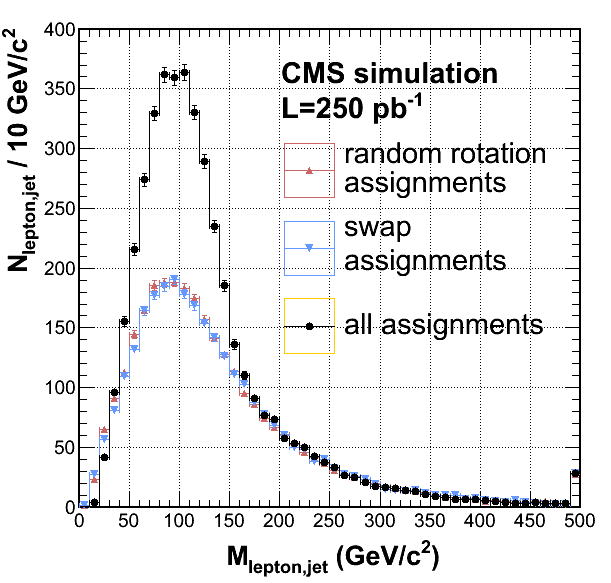} 
\caption{Invariant mass of the lepton-jet pairs for the whole data sample.}
\label{fig:Mlj_recodata}
\end{figure}

Despite small contributions from other background processes there is a non-negligible 
probability that at least one jet from a $t \bar{t}$ decay is either missed because it was not reconstructed or because it 
did not pass the jet selection criteria, and another jet is chosen instead (such as, for example, jets from ISR/FSR). 
This will be referred to as ``jet misassignment'' and an estimate of the jet misassignment level has to be
made from data. The estimate is done in terms of probability weights $\alpha_{i}$, where $i=0,1,2$ is the 
number of jets from top decays correctly reconstructed and selected.
The selected events are a combination of three different categories: 
\begin{itemize}
\item events with no jet selected from the top decays, 
weighted by $\alpha_{0}$ (background-dominated);
\item events with only one jet correctly assigned to the top decay, 
weighted by $\alpha_{1}$ (combination of signal and background);
\item events with two jets correctly assigned to the top decays, 
weighted $\alpha_{2}$ (signal-dominated)  .
\end{itemize}
In first approximation the weights $\alpha_i$ can be parameterized in terms 
of a binomial combination of $\alpha$, the probability
of correctly assigning individual jets.
The value of $\alpha$ can be estimated using the kinematic properties of the events directly from data.
A correlation can be sought in the lepton-jet pairs originating from the same top quark decay~\cite{RKELLIS}
and it is possible to show that no pair with $M_{l,b}>M_{l,b}^{max}\equiv\sqrt{m_t^2-m_W^2}$~=~156~GeV/c$^2$ 
should be observed (spectrum endpoint).
Two methods are proposed to emulate the invariant mass distribution of the misassigned jets:
``swapping'' the jet in the assigned lepton-jet pair, with a jet from a different event, 
or ``randomly rotating'' the momentum vector of the selected leptons.
As the ``random rotation'' and ``swap'' methods yield similar results, the average value is used
to model the invariant mass distribution of the background jets.
The distribution of the ``swapped'' and ``randomly rotated'' pairs, 
normalized to fit the high-end part of the distribution, is superimposed. 
The two background models provide a good estimate of the fraction of 
misassigned pairs with 
$M_{lepton,jet}>$~190~GeV/c$^2$ 
(Fig.~\ref{fig:Mlj_recodata}).
The normalization factor applied to the distribution of the swapped (randomly rotated) 
pairs is related to the misassignment fraction, $1-\alpha$ \cite{PASdilep}.
\subsection{Measurements by fitting the $n_{btag}$ distribution}
\label{subsec:fitDilep}
The following subsections describe the measurement of $R$ and of the b-tagging 
efficiency respectively; for both the measurements a fit of the $n_{btag}$ 
distribution is performed 
with the function in Eq.\ref{eq:epsilon} 
as in the semi-leptonic analysis,
 but here it depends also on $\alpha$ (besides $R$, $\epsilon_b$, $\epsilon_q$). 
In both the studies the $\alpha_2=\alpha^2$ parameter is fixed to the value obtained
 by data as explained above, $\alpha_0$ is a free parameter and $\alpha_1$ is obtained 
from the normalization ($\alpha_1$=1-$\alpha_0$-$\alpha_2$).
The value obtained for $\alpha_2$ 
is $\alpha_2=0.67 \pm 0.07$ (stat)$\pm 0.03$ (syst), while the one obtained by using the MC truth is $0.63 \pm 0.02$. 
$\epsilon_q$ is fixed to the value obtained by other data 
driven methods \cite{BTV_07_002} and the other parameters are fixed or free depending on the study.    
\subsubsection{Measurement of $\boldmath R$}
\label{subsub:rmeasurement}
\begin{figure}[h]
\centering
\includegraphics[height=80mm]{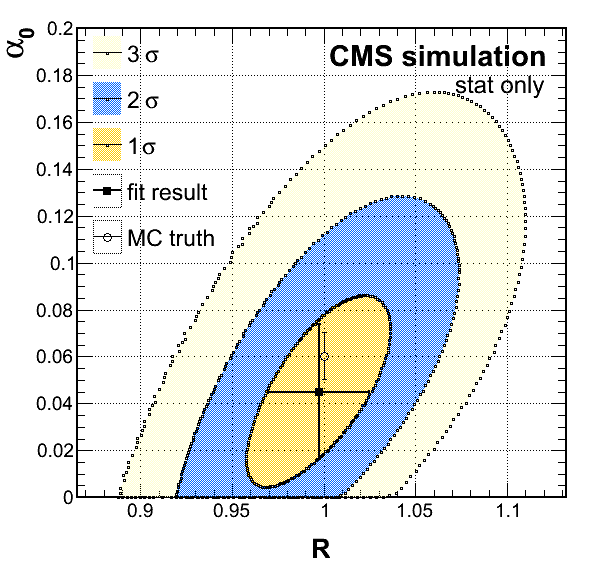} 
\caption{
Fit to $R$ and $\alpha_0$, only statistical uncertainties are shown.}
\label{fig:Rfit}
\end{figure}

In order to measure $R$, $\epsilon_b$ was fixed; the results for $R_{gen}=1$, for the two b-tagging algorithms ($JP$ and $TC$), each for three working points, are shown in Tab.\ref{tab:RFitTable}
\begin{table}[!h]
  \caption{$R$ fit results for an integrated luminosity of ${\mit L}$~=~250~pb$^{-1}$. 
            Statistical uncertainties from the fit and from MC truth are included.}
  \label{tab:RFitTable}
  \begin{center}
    \begin{tabular}{l|ccc}
      \hline
      $b$-tagging algorithm & \multicolumn{3}{c}{Working point} \\ \cline{2-4}
      & loose & medium & tight \\ \hline\hline
      Jet Probability  & 1.01 $\pm$ 0.02 & 1.00 $\pm$ 0.02 & 0.97 $\pm$ 0.03 \\ \hline
      Track Counting   & 1.00 $\pm$ 0.02 & 0.99 $\pm$ 0.02 & 1.04 $\pm$ 0.03 \\ \hline
    \end{tabular}
  \end{center}
\end{table}
Figure~\ref{fig:Rfit} shows the results obtained by fitting $R$
and $\alpha_0$ using jets tagged with the JP loose point. 
Different subsamples with forced decays, are weighted similarly to the semi-leptonic study 
(Sec.\ref{subsec:fitSemil}) in order to give different $R_{gen}$ values.
All backgrounds are included. 
The b-tag multiplicity distributions obtained this way
are sampled according to the expected number of events and fit to determine $R$.
The statistical uncertainty of each fit result is then determined from the width 
of the distribution of $R_{generated}-R_{measured}$.
The systematic uncertainties are dominated by the uncertainty on the $b$-tagging efficiency. The total uncertainty is $\sigma_{R}(stat+sys)=0.09$ with 250 pb$^{-1}$.
\subsubsection{Measurement of $\boldmath b$-tagging efficiency}
\label{subsub:effbmeasurement}
\begin{figure}[hbt]
\centering
\includegraphics[height=80mm]{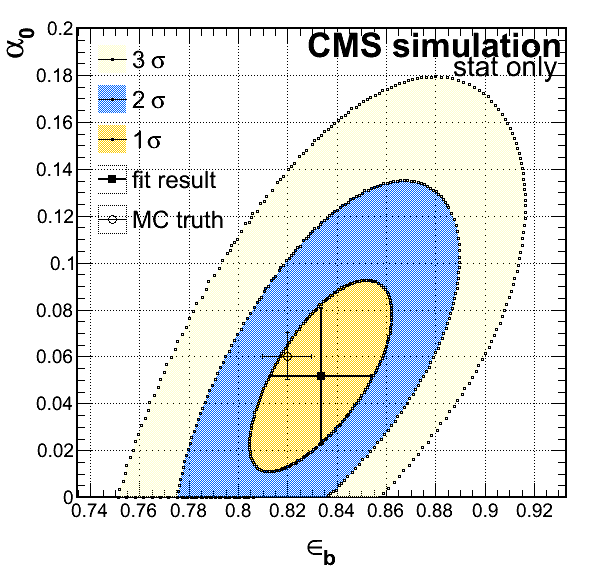} 
\caption{Contour plot of the fit to $b$-tagging efficiency and $\alpha_0$. $R=1$ fixed and
      $\alpha$ is fixed to the value estimated with the swap method.}
\label{fig:BtagFit}
\end{figure}

Here $R=1$ is fixed and the $b$-tagging efficiency, 
$\varepsilon_b$
is measured. 
The constraint $0\leq \varepsilon_b\leq 1$ is used in the fit.
The results are shown in Tab.~\ref{tab:BtagEffFitTable}.
\begin{table}[h]
  {
    \caption{Fit to $b$-tagging. $R=1$ fixed and
      $\alpha$ is fixed to the value estimated with the swap method.
      Statistical uncertainties from the fit and from MC truth are included.}
    \label{tab:BtagEffFitTable}
    \begin{center}
      \begin{tabular}{cccc} \hline
	algorithm       & working point & $\varepsilon_{b}$~(MC truth) & $\varepsilon_{b}$ \\ \hline \hline
	                & loose  & $0.82 \pm 0.01 $ & $0.81 \pm 0.02$ \\ \cline{2-4}
	Jet Probability & medium & $0.63 \pm 0.01 $ & $0.63 \pm 0.02$ \\ \cline{2-4}
	                & tight  & $0.41 \pm 0.01 $ & $0.41 \pm 0.02$ \\ \hline
	               & loose  & $0.80 \pm 0.01 $ & $0.82 \pm 0.02$ \\ \cline{2-4}
	Track Counting & medium & $0.65 \pm 0.01 $ & $0.65 \pm 0.02$ \\ \cline{2-4}
	               & tight  & $0.40 \pm 0.01 $ & $0.41 \pm 0.02$ \\ \hline
      \end{tabular}
    \end{center}
  }
\end{table}
A simultaneous fit to the $b$-tagging efficiency and $\alpha_0$ yields the 2-dimensional
distribution shown in Figure~\ref{fig:BtagFit}.
The total systematic uncertainty is $4\%$ and is due to the uncertainty on $\alpha$.
The uncertainty is estimated by repeating the fit procedure after displacing each 
parameter by positive/negative values from a Gaussian distribution centered at zero 
with a width given by the corresponding uncertainty of the parameter.
The sensitivity of the $\varepsilon_b$ measurement is about $\pm 0.02$ when $R$ is varied by $5\%$.
The fitting model is derived for $t \bar{t}$ events and the bias is estimated to be small,
given that the background events are only $~10\%$ of the total sample. The good agreement (within uncertainties)
of the fit results with the MC truth values 
justifies this assumption.
The uncertainty due to different ISR/FSR content in the final sample is expected to be small ($<1\%$).
\section{Conclusions}
Two studies of feasibility of the $R$ measurement was presented, 
one by using selected semi-leptonic $t \bar{t}$ events and the other 
by using selected di-leptonic $t \bar{t}$ events in the $e \mu$ channel.
The expected uncertainties, for the semi-leptonic channel with $L=1$ fb$^{-1}$, 
are $\sigma_R(stat) = 0.12$ and $\sigma_R(sys) = 0.11$. 
For the di-leptonic channel, with $L=250$ pb$^{-1}$, the expected uncertainty is 
$\sigma_R(stat+sys) = 0.09$. Furthermore, in the dileponic channel a study 
on the $\epsilon_b$ measurement, 
fixing $R$ to the SM value, has been performed. The expected uncertainties are:
$\sigma_{\epsilon_b}(stat)=0.02$ $\sigma_{\epsilon_b}(sys) \sim 0.04$.
Both the studies use data driven methods to subtract the background contribution.


\begin{thebibliography}{99} 
\bibitem{Alwall:2006bx}
J.~Awall et al
{\em Eur. Phys. J.} C49 791-801(2007).

\bibitem{Acosta:2005hr}
D.~Acosta et al., CDF Collaboration,
{\em Phys. Rev. Lett.} 95 102002(2005)[hep-ex/0505091].

\bibitem{Abazov:2008yn}
V.~M.~Abazov et al., D0 Collaboration, 
{\em Phys. Rev. Lett.} 100,192003(2008).

\bibitem{Amsler:2008zzb}
C.~Amsler et al.,
{\em Phys. Lett.} B667,1(2008).

\bibitem{Abazov:2009ii}  
V.~M.~Abazov  et al. [The D0 Collaboration],  
arXiv:0903.0850[hep-ex].

\bibitem{Aaltonen:2009jj}
T.~Aaltonen et al. [The CDF collaboration],
arXiv:0903.0885 [hep-ex].
\bibitem{cms}
CMS Collaboration, 
{\em The CMS experiment at the CERN LHC} JINST 3:S08004,2008.
\bibitem{PASsemil}
CMS Physics Analysis Summary TOP-09-007 http://cdsweb.cern.ch/record/1194522?ln=en
\bibitem{PASdilep}
CMS Physics Analysis Summary TOP-09-001 http://cdsweb.cern.ch
\bibitem{PTDRVOL1}{\it CMS Physics Technical Design Report Vol.1},CERN/LHCC 2006-001.

\bibitem{BTV_07_001} CMS PAS BTV-07-001.

\bibitem{NotaCMS_B}CERN-CMS-NOTE-2006-019.


\bibitem{RKELLIS} Ellis,~R.K. et al., Cambridge Monographs on Part.~Phys., Nucl.~Phys. and Cosmology.

\bibitem{BTV_07_002}CMS PAS BTV-07-002.

\end{thebibliography}

\end{document}